\documentstyle[prb,aps,epsf,twocolumn]{revtex}

\begin{document}
\draft

\title{Melting behavior of ultrathin titanium nanowires}

\author{Baolin Wang $^{a,c}$, Guanghou Wang $^a$, Jijun Zhao $^b$}

\address{$^a$ National Laboratory of Solid State Microstructures and Department of Physics, Nanjing University, Nanjing 210093, China \\
$^b$ Department of Physics and Astronomy, University of North Carolina at Chapel Hill, Chapel Hill, NC 27599, USA.\\
$^c$ Department of Physics, Huaiyin Teachers College, Jiangsu, 223001, China}
\maketitle

\begin{abstract}

The thermal stability and melting behavior of ultrathin titanium nanowires with multi-shell cylindrical structures are studied using molecular dynamic simulation. The melting temperatures of titanium nanowires show remarkable dependence on wire sizes and structures. For the nanowire thinner than 1.2 nm, there is no clear characteristic of first-order phase transition during the melting, implying a coexistence of solid and liquid phases due to finite size effect. An interesting structural transformation from helical multi-shell cylindrical to bulk-like rectangular is observed in the melting process of a thicker hexagonal nanowire with 1.7 nm diameter.

\end{abstract}


\vspace{-0.2in}
Metal nanowires are current focus of intensive research due to both the fundamental interests of low-dimensional physics and the potential applications in nanoscale materials and devices. Experimentally, stable ultrathin metal nanowires with diameter down to several nanometers and sufficient length have been fabricated by various methods \cite{1,2,3,4,5,6}. For example, Ti nanowires of a few nm in width were produced by Ar$^{+}$ ions irradiation on Ti thin layer by using carbon nanotube as a mask \cite{1}. Takayanagi's group has successfully produced suspended stable gold wires \cite{2,3}. Novel helical multi-shell structures were observed in those ultrathin gold nanowires \cite{3}. As compared with bulk materials and atomic clusters, one-dimensional (1-D) metal nanowire have both the 1-D periodicity along wire axis similar to bulk solids and the larger surface to volume ratio like clusters. Therefore, the metal nanowires are expect to exhibit some unique features different from either bulk solids or nanoclusters. 

On theoretical side, the structures and properties of free-standing ultrathin metal nanowires have been studied by several groups \cite{7,8,9,10,11,12,13,14,15,16,17}. Based on genetic algorithm simulations with empirical potentials, we have systematically studied the helical multi-shell structures in Au \cite{14}, Ti \cite{15}, Zr \cite{16} nanowires. The corresponding vibrational, electronic, and transport properties were also discussed. However, little is known about the thermal stability and melting behavior of the 1-D metal nanowires, especially for those transition metal nanowires with novel cylindrical multi-shell structures. Thus, it is important to clarify the relation between the melting behavior and the size or geometry of these ultrathin metal nanowires. The finite size effects \cite{18,19,20,21,22}, e.g., solid-liquid coexistence, on the phase transition of such quasi-1D systems are also interesting. For thicker crystalline lead nanowires, surface melting phenomena associated with formation of a thin skin of highly mobile surface atoms were observed \cite{7}. More recently, the multi-shell gold nanowires and their melting behaviors have been studied by MD simulations \cite{13}. In our previous work, the atomic structures of titanium nanowires with diameter from 0.7 to 1.7 nm have been optimized and helical multi-walled cylindrical structures have been obtained \cite{15}. In this paper, we report molecular dynamic simulations on the thermodynamic properties of the titanium nanowires. The effect of nanowire size and geometry on the melting behavior will be discussed. 

In our simulations, Ti nanowires of sufficient length are modeled by supercell with 1-D periodical boundary condition along the wire axis direction. As a reasonable compromise between keeping the helicity in $z$ axis direction and avoiding the nanowire breaking into clusters upon relaxation, the length of the supercell is chosen to be 1.256 nm.  The interaction between titanium atoms is described by a well-fitted tight-binding many body potential \cite{23}. To simulate the melting behavior of nanowires, we employ the constant temperature molecular dynamics (MD) method by Hoover \cite{24}, which have been extensively used in our previous works \cite{25,26,27}. The MD time step is chosen as 2.15 fs. At each temperature, the initial 10$^5$ MD steps are used to bring the system into equilibrium. Then we monitor the internal energy $E$, root-mean-square (rms) fluctuation of the interatomic distances $\delta $, and heat capacity $C_v$ from the thermal statistical averages in the equilibrium canonical ensemble. At each temperature, 10$^6$ MD steps are performed to record the thermodynamic average of these physical properties. The constant-temperature MD simulations start from a low temperature (400 K). The temperature gradually increases towards high temperature (1400 K) by 50 K per step.

We first perform comprehensive genetic algorithm simulations \cite{14,15} on the equilibrium structures of titanium nanowires. Several representative structures are presented in Fig.1. These stable multi-shell structures are composed by coaxial cylindrical shells, which have been theoretically predicted for Al, Pb, Zr, Cu, and Au nanowires \cite{8,12,13,14,16} and experimentally observed in Au nanowires \cite{3}. We use index $n$-$n^{\prime }$-$n^{{\prime}{\prime}}$-$n^{{\prime}{\prime}{\prime}}$ \cite{3,14} to describe the nanowire consisting of coaxial shells with $n$, $n^{\prime}$, $n^{{\prime}{\prime}}$, $n^{{\prime}{\prime}{\prime}}$ helical atomic rows ($n>n^{\prime}>n^{{\prime}{\prime}}>n^{{\prime}{\prime}{\prime}}$) each. Thus, the nanowire structures in Fig.1 can be defined as 8-2, 9-3, 9-4, 5-1, 6-1, 12-6-1, and 17-12-6-1 respectively. The structure of the thinnest 5-1 wire is centered pentagonal, with the outer shell containing five strands and the inner shell as a single atomic row. The 6-1, 12-6-1, and 17-12-6-1 wires in Fig.1 constitute a hexagonal growth sequence containing two-shell, three-shell, and four-shell with a central single atomic row. In stead of central atomic row based structures, the 8-2, 9-3, 9-4 wires are composed of eight or nine strands in the outer shell and two to four strands in the inner shell. Moreover, multi-shell structures such as 13-8-2, 14-9-3, 15-9-4 are obtained in our simulation. Our systematical illustration of various multi-shell growth sequence of the metal nanowire can be found elsewhere \cite{16}.

\begin{figure}
\vspace{-0.1in}
\centerline{
\epsfxsize=2.95in \epsfbox{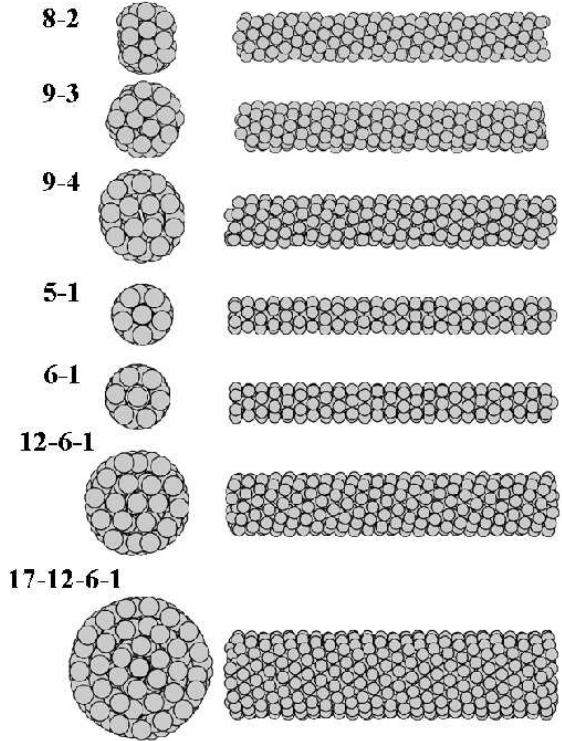}
}
\caption{Morphology of Ti nanowires with diameter from 0.75 to 1.7 nm. Top view (left) and side view (right) are presented. These multi-shell structures can be identified as 8-2, 9-3, 9-4, 5-1, 6-1, 12-6-1, and 17-12-6-1 respectively.}
\end{figure}

Starting from those optimized structures at low temperature, we simulate the melting behavior of the titanium nanowires using constant-temperature MD method. As representative examples, Fig.2 (a), (b), (c), (d), (e) present the temperature dependence of the internal energy $E$, mean square bond length fluctuation $\delta$, and thermal capacity $C_v$ for 5-1, 6-1, 9-4, 12-6-1, and 17-12-6-1 nanowires respectively. From those curves, we can define the overall melting temperature T$_m$ of the system as the temperature at half of saturated $\delta $ in liquid states \cite{??}. For the purpose of comparison, we have also simulated the melting of some titanium nanoclusters with equilibrium structures such as icosahedral Ti$_{55}$ \cite{27}. Table 1 summarizes the T$_m$ of various nanowires with different multi-shell structural pattern from our simulations. In general, the melting temperatures of titanium nanowires are much lower than the bulk value (1943 K) and increase with the nanowire size. Because of the 1-D periodicity and the well-reconstructed helical surface structures, the melting temperatures of nanowires are typically higher than those of clusters with comparable size (see Table I).

\begin{figure}
\vspace{-0.35in}
\vspace{0.0in}
\centerline{
\epsfxsize=2.25in \epsfbox{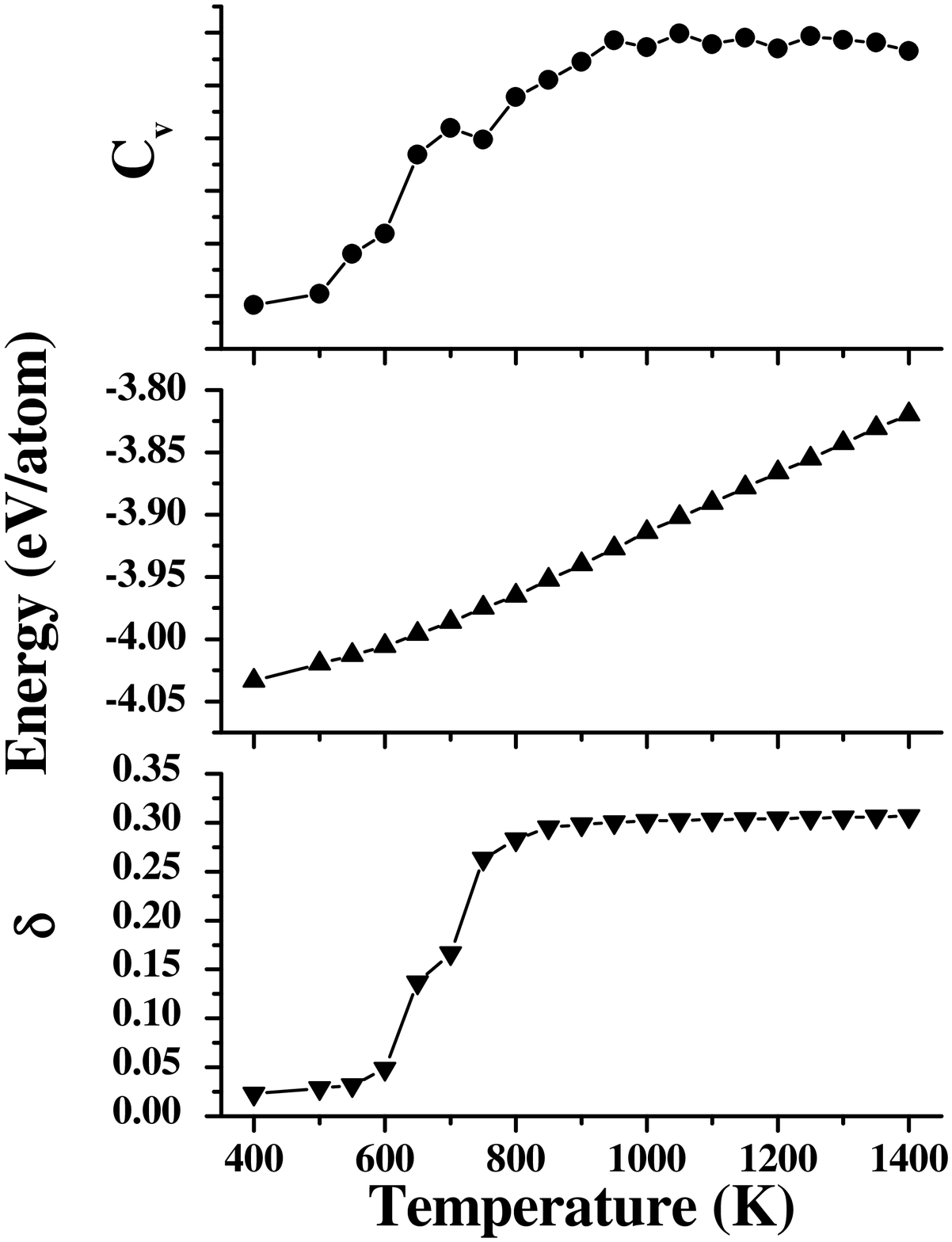}
}
\vspace{-0.15in}
\centerline{
\epsfxsize=2.25in \epsfbox{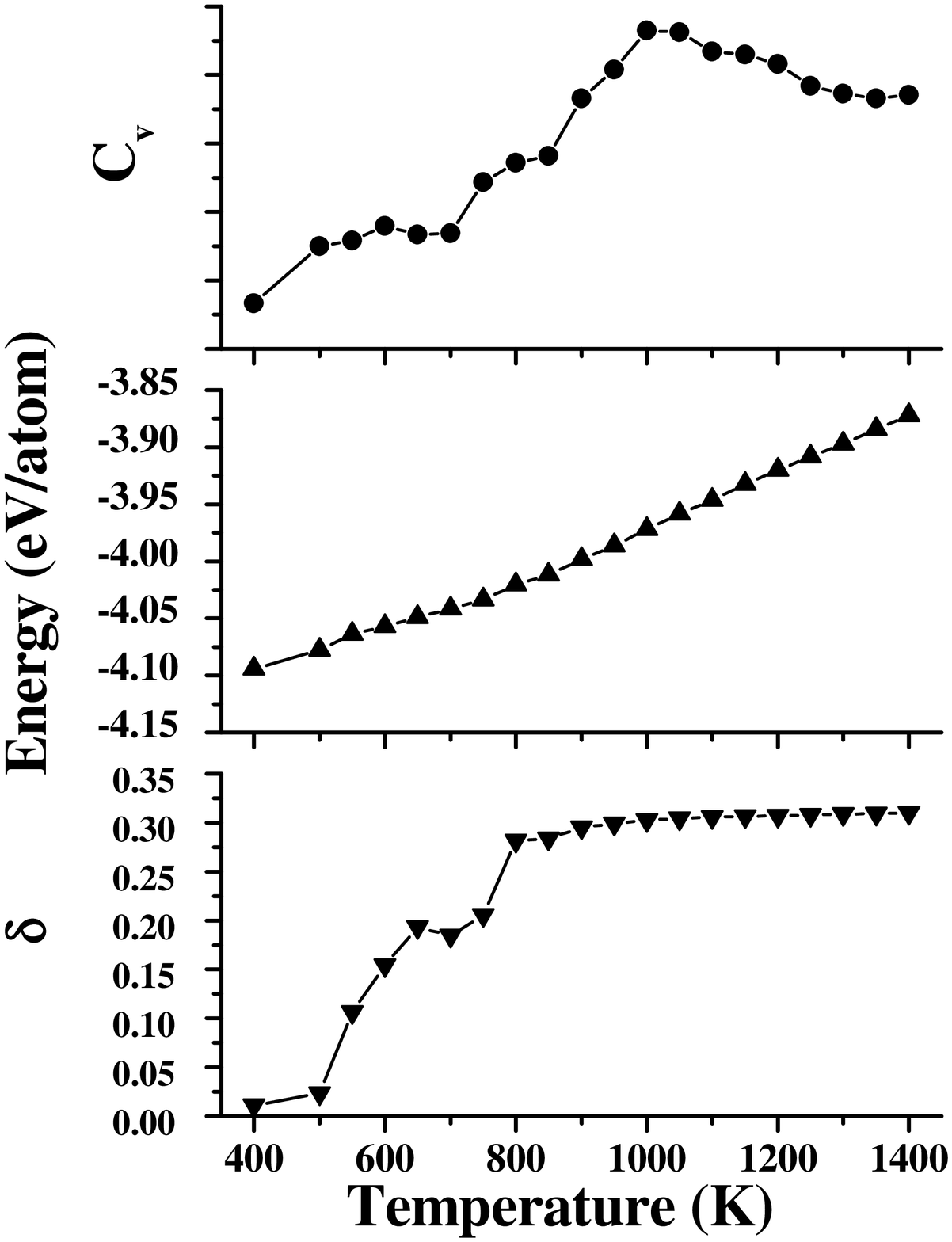}
}
\vspace{-0.15in}
\centerline{
\epsfxsize=2.25in \epsfbox{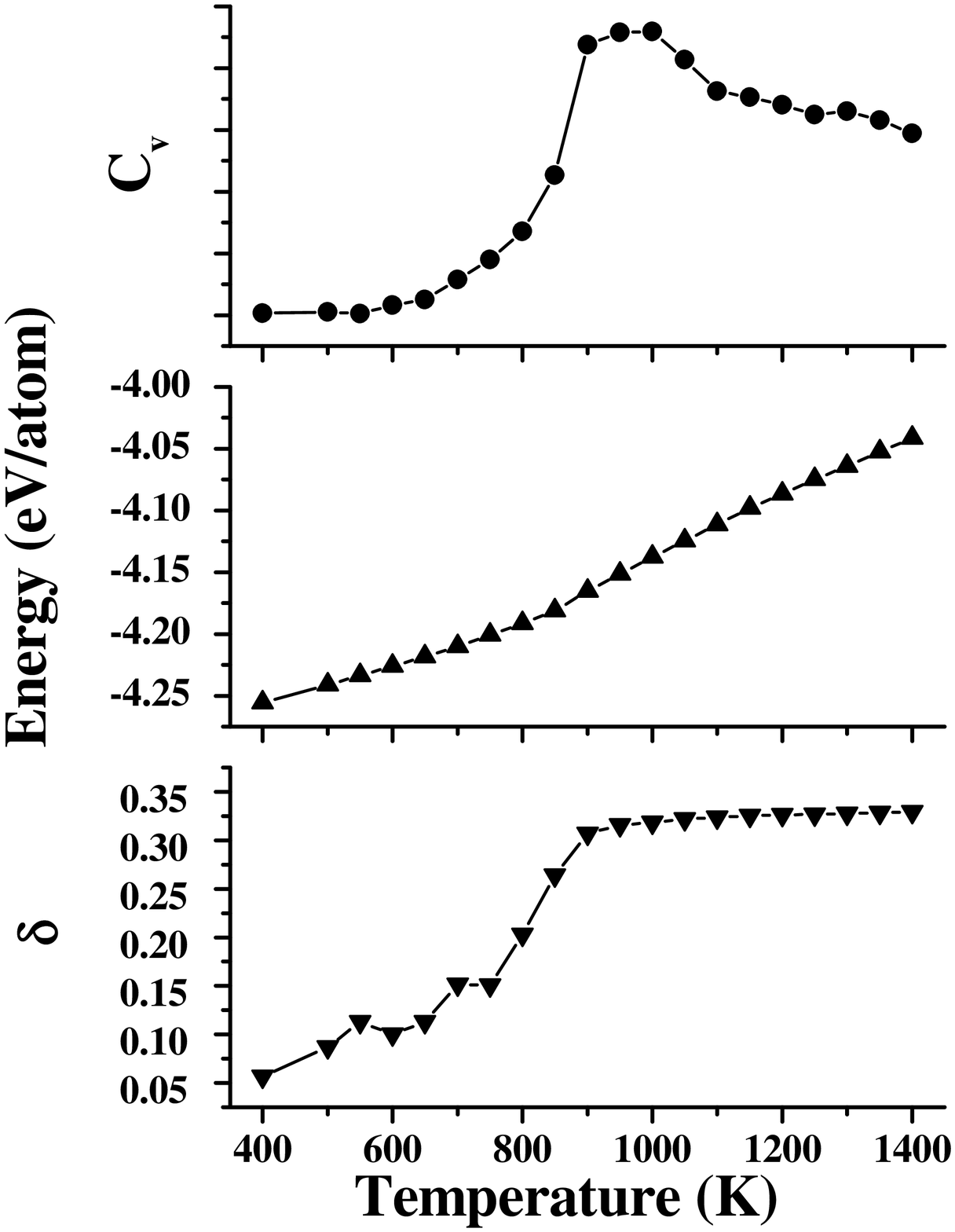}
}
\vspace{-0.15in}
\centerline{
\epsfxsize=2.25in \epsfbox{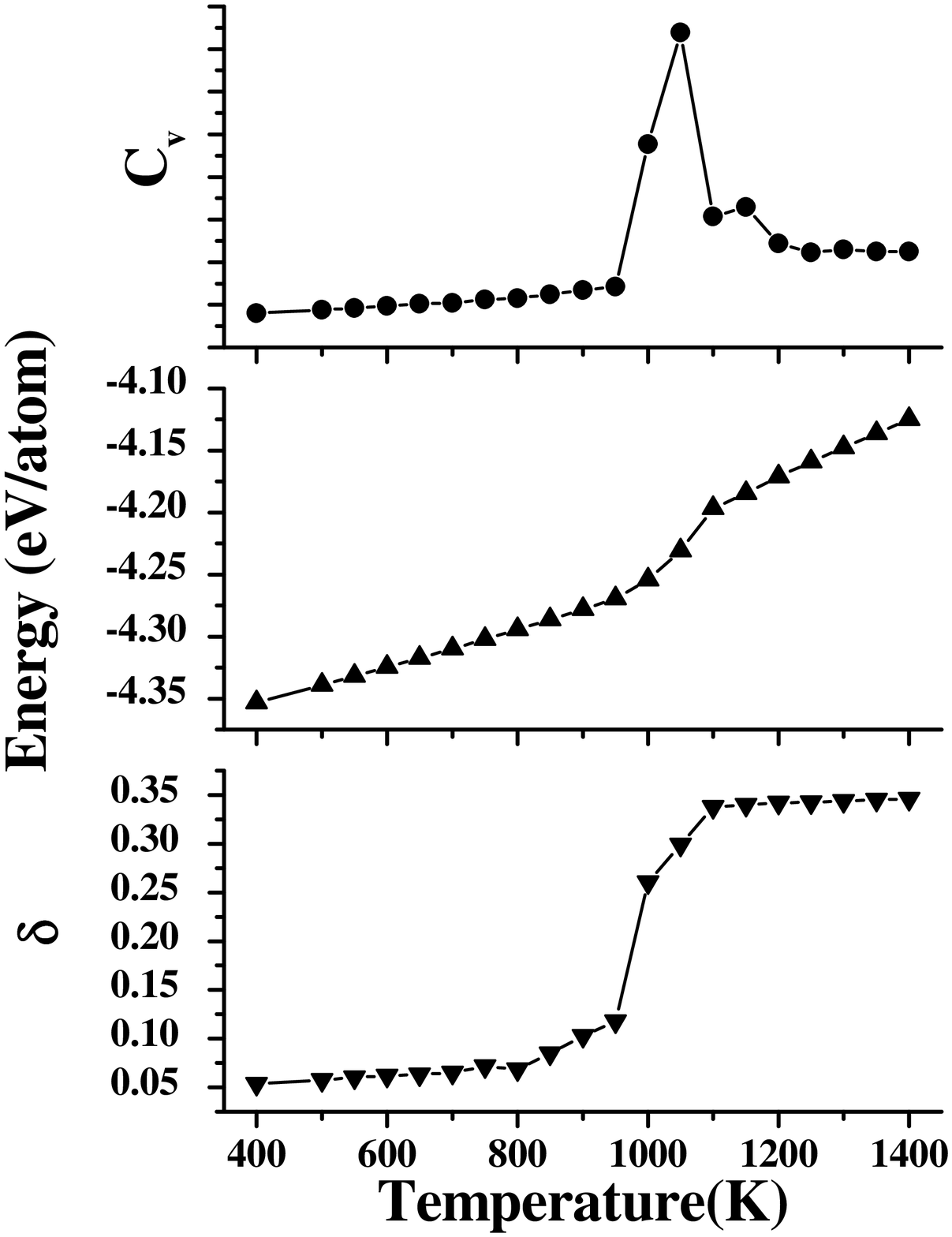}
}
\vspace{-0.15in}
\centerline{
\epsfxsize=2.25in \epsfbox{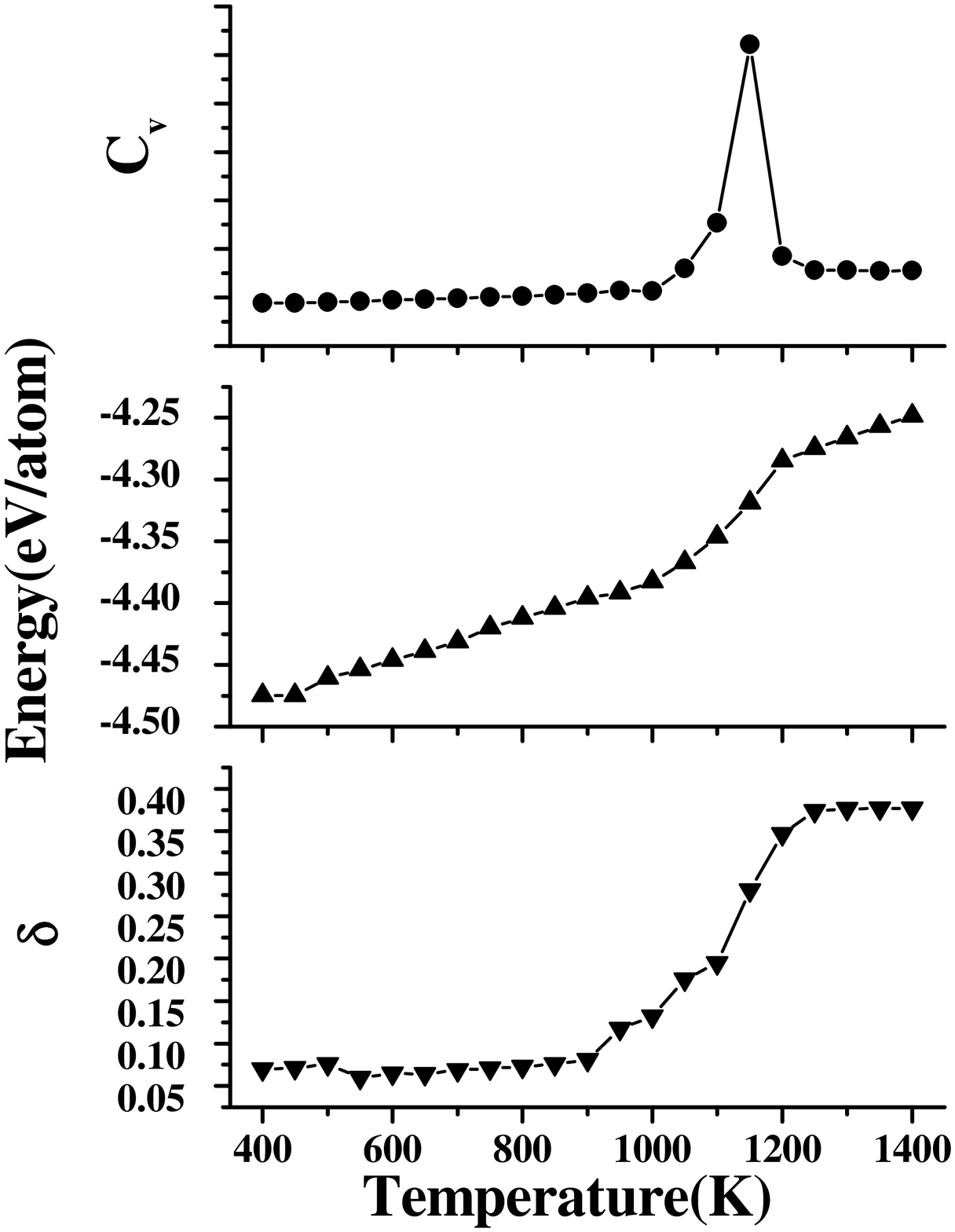}
}
\caption{The rms bond length fluctuation $\delta$, internal energy $E$, and heat capacity $C_v$ as functions of nanowire temperature for various nanowires. (a): 5-1; (b): 6-1; (c) 9-4; (d) 12-6-1; (e) 17-12-6-1.}
\end{figure}

\vspace{-0.1in}
\begin{table}
Table I. Overall melting temperatures T$_{m}$ for titanium nanowires and clusters. The diameter D of the nanowires or cluster are given.

\begin{tabular}{ccccccc}
             & 8-2 & 13-8-2 & 9-3 & 14-9-3 & 9-4 & 15-9-4 \\ \hline
 D (nm)      & 0.93& 1.43   & 1.04& 1.52   & 1.07& 1.60   \\ 
T$_{m}$ (K)  & 800 & 1050   & 850 & 1000   &  850& 950   \\
\end{tabular}

\begin{tabular}{ccccccc}
             &  5-1 & 10-5-1 & 6-1 & 12-6-1 & 17-12-6-1 & Ti$_{55}$ \\ \hline
 D (nm)      & 0.75 & 1.18   & 0.81& 1.28   & 1.71      &  1.23     \\ 
T$_{m}$ (K)  &  800 & 900    &  700 & 1000  & 1150      &   850     \\
\end{tabular}
\end{table}

To explore the size dependence of nanowire melting temperature, we plot the overall melting temperature $T_m$ versus wire diameter (1/D) in Fig.3. It is found that the $T_m$ for the 6-1, 12-6-1, 17-12-6-1 wires under the same hexagonal growth sequence fit well to a linear dependence as:
$$
T_m=T_0-\eta /D
$$
where $T_0=$ 1542 K is the extrapolated infinite limit, $\eta= 682$ K ${\cdot}$ nm describes the linear dependence of $T_m$ with diameter $D$ (in unit of nm). Such $1/D$ dependence of melting temperature for nanowire is similar to the well-known size relationship for metal nanoclusters \cite{28,29,30}. As compared with the linear dependence in the hexagonal wires (6-1, 12-6-1, 17-12-6-1),  8-2 and 13-8-2 wires almost belong to the same size relationship while the $T_m$ of other nanowires deviate from such linear fit. These differences indicate that the atomic structures of nanowires play significant role in determining the melting behavior of nanowires. Among those wires, the $T_m$ of 5-1 wire is much higher than the 6-1 wires with comparable size. On  the other hand, the nanowires with three or four atomic strands in the internal shell like 9-3, 14-9-3, 9-4, 15-9-4, have lower melting temperature than the nanowires with one atomic strand in the center, e.g., 12-6-1. This effect might be understood by the relatively looser internal structures of these 9-3 or 9-4 based nanowires (see Fig.1). Therefore, we suggest that the multi-shell metal nanowires with tightly internal structures should have higher thermal stability. 

\begin{figure}
\vspace{0.5in}
\centerline{
\epsfxsize=2.8in \epsfbox{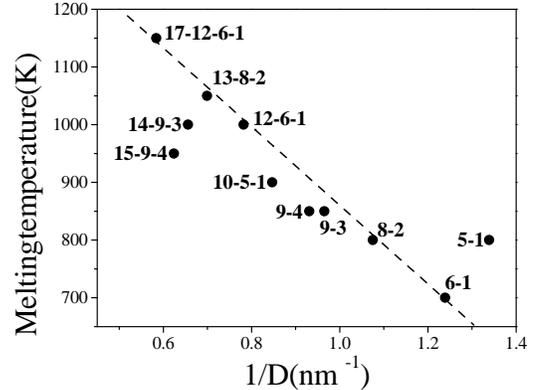}
}
\vspace{-0.65in}
\caption{Overall melting temperature $T_m$ (K) vs. nanowire diameters 1/D (1/nm) for various titanium nanowires with multi-shell structures. Dashed line is linear fit for hexagonal wires, i.e., 6-1, 12-6-1, 17-12-6-1.}
\end{figure}

We now discuss the nanowire melting behavior by examining the temperature dependent curve of internal energy $E$, bond fluctuation ($\delta$) and thermal capacity ($C_v$) with different size and geometries. As shown in Fig.2(a-e), the melting of 5-1, 6-1 and 9-4 wires all start from about $500-600$ K. Their difference in the internal structures are only reflected in the slightly different variations of rms bond length fluctuation $\delta$ with temperature. Similar effect has been also found in other thinner nanowires like 8-2, 9-3. For all the thinner wires (e.g., 5-1, 6-1, 9-4 in Fig.(a-c)), we haven't seen clear peak in the $C_v$ curve in Fig.2, which is a characteristic for the first-order solid to liquid phase transition. Accordingly, the increase of internal energies is also relatively smooth in these small nanowires. Only the dramatic jump in bond length fluctuation $\delta$ with increasing temperature demonstrates the occurrence of melting. This behavior is similar to that in small clusters \cite{21}. It can be understood by finite size effect in first-order phase transition \cite{18,19}, which leads to the coexistence of solid and liquid states in a rather broad temperature region \cite{20,21,22}. On the other hand, the thicker wires, such as 12-6-1 and 17-12-6-1, show clear characteristics of the solid to liquid phase transition (see Fig.2(d,e)). In particular, there are sharp peaks in the thermal capacity curves at $1050$ K and $1100$ K for 12-6-1 and 17-12-6-1 wires respectively. 

\begin{figure}
\vspace{-0.05in}
\centerline{
\epsfxsize=3.1in \epsfbox{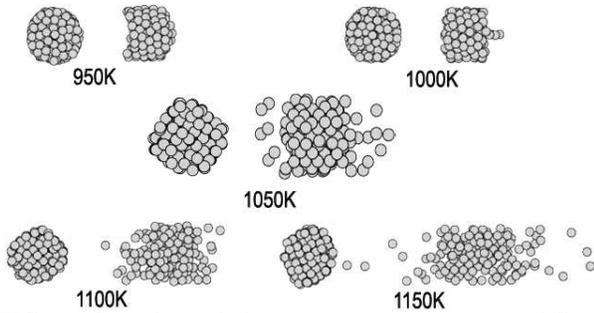}
}
\caption{Snapshot of the 17-12-6-1 nanowire at different temperatures during the melting process. A structural transformation is clearly seen at 1050 K.}
\end{figure}

To further understand the melting process in the titanium nanowires, we have traced the structural change of the 17-12-6-1 wire with the temperatures in the melting region. Fig.4 gives the snapshot of the nanowire at different temperatures. One can see that the titanium atoms only weakly oscillate at their equilibrium positions before the overall melting ($950-1000$ K). Although there are small deformation on the sharp of nanowire and the diffusion of interior atoms, the top view of the nanowire cross-section still shows hexagonal structure with distinct atomic shells. It is interesting to find a structural transformation of the nanowire at $1050$ K. In additional to the thermal enhanced 1-D diffusion along wire axis, the atomic arrangements of the nanowire have transformed from cylindrical helical to bulk-like rectangular. In the further melting process, the structures of nanowire remains bulk-like. The structural transition from cylindrical helical to bulk-like rectangular has also been observed in the size evolution of gold nanowire optimized structures in our previous simulation \cite{14}. 

From the above results and discussions, we make the following conclusions. (1) The thermal stability of titanium nanowires significantly dependent on the structures and sizes of the wire. (2) The melting temperature of titanium nanowire is lower than the bulk's, but higher than titanium cluster's with comparable size. Linear dependence of melting temperature upon inverse of nanowire diameter ($1/D$) is obtained. (3) The coexistence mechanism of the solid state and the liquid state due to finite size effect plays a significant role in the melting process of the thinner nanowires ($D < 1.2 $nm). (4) For the thicker nanowires, i.e., $D\sim 1.7$ nm, we find the structural transformation from the helical multi-shell cylindrical to the bulk-like during the melting process. 

The authors thank financial support from National Nature Science Foundation of China (No.29890210).

\end{document}